\documentclass[12pt]{article}
\usepackage{graphicx,epsfig}
\usepackage{cite}

\begin{document}

\title{Z-lineshape versus 4th generation masses.}
\author{S. S. Bulanov \footnote{bulanov@heron.itep.ru}, V. A. Novikov
\footnote{n2ovikov@heron.itep.ru}, L. B. Okun \footnote{okun@heron.itep.ru}, \\
{\small ITEP, Moscow, Russia} \\
A. N. Rozanov \footnote{rozanov@cppm.in2p3.fr},\\
{\small CPPM, IN2P3, CNRS, Univ. Mediteranee, Marseilles, France} \\
{\small and ITEP, Moscow, Russia}\\
M. I. Vysotsky \footnote{vysotsky@heron.itep.ru},\\
{\small ITEP, Moscow, Russia.}
}
\date{}
\maketitle

\begin{abstract}
The dependence of the Z-resonance shape on the location of the
threshold of the $N\bar{N}$ production ($N$ is the 4th generation
neutrino) is analyzed. The bounds on the existence of 4th
generation are derived from the comparison of the theoretical
expression for the Z-lineshape with the experimental data. The 4th
generation is excluded at 95\% C. L. for $m_N<46.7\pm 0.2$ GeV.
\end{abstract}

\section{Introduction}
The straightforward generalization of the Standard Model through
the inclusion of extra chiral generations of heavy leptons (N, E)
and quarks (U, D) was studied in a number of papers 
\cite{4th_1,4th_2,4th_3,4th_4,4th_5,4th_6,4th_7,4th_8,4th_9,turk}.
In \cite{4th_1,4th_2,4th_3} the analysis of deviations from the Standard Model 
due to 4th generation contribution was carried out in terms of $S$, $T$ and 
$U$ parameters for 4th generation particles being much heavier than $m_Z$. The 
case of new light physics was investigated in \cite{4th_4,4th_5}, using 
modified $S$, $T$ and $U$ parameters in order to take into account the effects 
of relatively light new particles. Particle and astroparticle implications of 
the 4th generation neutrinos were studied in \cite{4th_7}. A more thourough 
investigation of the properties of the 4th generation in the framework of GUT 
models was carried out in \cite{4th_8}. It was shown in \cite{4th_9} that 
unifying spins and charges in the framework of SO(1,13) group one gets four 
families of leptons and quarks. Possible manifestations of 4th generation 
particles at hadron colliders were studied in \cite{turk}.

The bounds on the existence of the 4th generation from the
analysis of the electroweak data fit were obtained in
\cite{4th_6,1,2,4,5}. However, the dependence of Z-resonance shape
on the contribution of 4th generation and, in particular, on the
location of the threshold of $N\bar{N}$ production was not
considered in \cite{1,2,4,5}, because the results were obtained in
the Breit-Wigner approximation. This approximation is valid for
the thresholds of 4th generation particles production being far
from $m_Z$. If the threshold location approaches $m_Z$
($m_N\rightarrow m_Z/2$), than one gets a fast worsening of the
fit (see Fig. 4 of \cite{4}).  It is due to the fact that the
standard approach to the radiative corrections to the electroweak
observables, used in \cite{4th_6,1,2,4,5}, does not work in the
presence of the heavy neutrino (N) with $m_N-m_Z/2<\Gamma_Z$.
After Taylor expansion over $p^2-m_Z^2$ the expression for the
polarization operator tends to infinity for $m_N\rightarrow
m_Z/2$, because the point, at which the Taylor expansion  is
performed, becomes the branch point of the polarization operator.

According to the results of \cite{1,2,4,5} the best fit
corresponds to $m_N\approx 50$ GeV, that is why careful analysis
of the region $m_N\approx m_Z/2$ is undertaken in this paper. We
study the dependence of the Z-lineshape on the location of the
threshold of $N\bar{N}$ production. We analyze the energy
dependence of $e^+e^-\rightarrow Z\rightarrow hadrons$ cross
section near Z-resonance and find that it exhibits a
characteristic behavior near the threshold, a casp \footnote{Such
behaviour of the cross section in quantum mechanics was discovered
by Wigner, Baz and Breit and is discussed in textbook
\cite{landau}. In particle physics analogous phenomenon was
considered in \cite{okun}. Unlike cases analyzed previously,
Z-boson physics is purely perturbative, allowing to get explicit
formulae for cross section. However, being perturbative the
variation of cross section because of the casp are small.
Nevertheless, high precision of experimental data on Z production
allow us to bound N mass from below.}. The casp is caused by the
square root ($\sqrt{s-4m_N^2}$), appearing in the contribution of
4th generation neutrino to the polarization operator of Z-boson.
The form of the casp is determined by the location of the
threshold with respect to $m_Z$.

Then we compare the theoretical expression for the Z-lineshape
with the experimental data , presented in \cite{9}, using the
ZFITTER \cite{10} in order to take into account the
electromagnetic corrections, and find that the 4th generation is
excluded at 95\% C. L. for $m_N<46.7\pm0.2$ GeV. This bound
depends on the masses of the charged 4th generation particles and
on the mass of the higgs. Using the results of \cite{5}, we fix
the mass of the charged lepton (E) and take into account that the
splitting of quark masses and higgs mass are not independent. This
leaves us with one free parameter -- the splitting of quark
masses,  which we vary from 0 to 50 GeV. This variation is the
source of the theoretical uncertainty in the bound on N mass  as
well as the uncertainties of the input parameters of ZFITTER.
Note, that the $\chi^2/n_{d.o.f.}$ for the Z-lineshape with 4th
generation is even better for certain region of mass values then
the $\chi^2/n_{d.o.f.}$ for the SM prediction.

The paper is organized as follows. In section 2 we discuss the
general behaviour of the cross-sections near threshold. In section
3 the exact formulae for the contributions of 4th generation
particles to the cross section of $e^+e^-\rightarrow hadrons$ are
presented. We study the behavior of the $e^+e^-\rightarrow
hadrons$ cross section near the threshold of $N\bar{N}$ production
in section 4. Using the result of section 3 we compare our
prediction for the Z-lineshape in the presence of 4th generation
with the experimental data in section 5. The conclusions are
presented in section 6.

\section{The cross-sections near threshold.}

Let's consider the interaction of two particles A and B, which
form some resonance R, which in its turn decays into a system of
particles f (see Fig. 1). The behavior of the cross section of
this process near R-peak can be calculated in the general case,
regardless of the exact form of interaction. We need only the
partial width of particle R. The cross section near the resonance
is described by the Breit-Wigner formula \cite{10}:
\begin{equation}
\sigma=\frac{4\pi s^2}{I^2}\frac{2S_R+1}{(2S_A+1)(2S_B+1)}
\frac{\Gamma_{R\rightarrow A+B}\Gamma_{R\rightarrow
f}}{(s-M^2)^2+\Gamma^2 s^2/M^2}\frac{s}{M^2}, \label{BW}
\end{equation}
where $s=(p_A+p_B)^2$, $M$ and $\Gamma$ are mass and total width
of R correspondingly, $I=1/2\sqrt{(s-(m_A+m_B)^2)(s-(m_A-m_B)^2)}$. $S_R$,
$S_A$ and $S_B$ are spins of particles R, A and B correspondingly.

\begin{figure}[h!]
\begin{tabular}{ccc}
\phantom{aaaaaaaaaaaaaa} &\epsfxsize6cm\epsffile{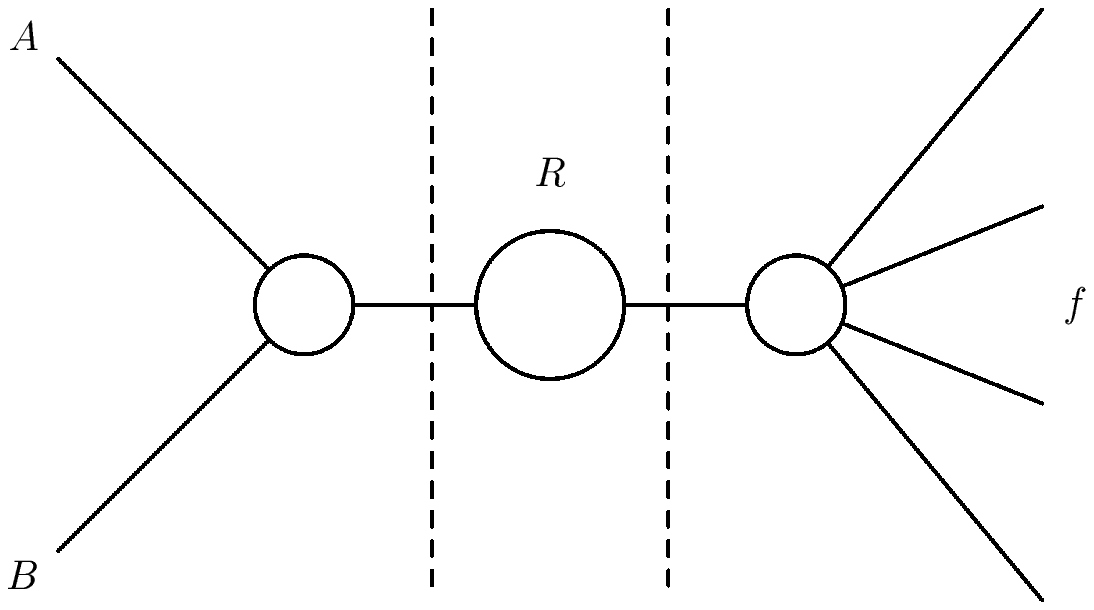}&
\end{tabular}
\caption{The reaction $A+B\rightarrow R\rightarrow f$.}
\end{figure}

Let's consider the case when the reaction occurs not only near
resonance, but also near the threshold of $N\bar{N}$ production.
In order to study this effect we write explicitly the contribution
of the $N\bar{N}$ loop to the propagator of particle R:
\begin{equation}
\frac{1}{s-M^2+i\Gamma s/M+\Sigma^{(N)}_R(s)},
\end{equation}
the imaginary part of polarization operator $\Sigma^{(N)}_R(s)$ is
connected with the $R\rightarrow N\bar{N}$ decay probability by
the unitarity relation. The decay probability is proportional to
$\sqrt{s-4m_N^2}$, the factor that arises from the integration
over phase space. Then, if we rewrite the polarization operator as
$\Sigma^{(N)}_R(s)=a+ib\sqrt{s-4m_N^2}$, and expand the expression
for the propagator near $s=4m_N^2$, it will take the following
form
\begin{equation}
T_0+iT_1\sqrt{s-4m_N^2},
\end{equation}
where $T_0$ and $T_1$ are some functions of $a$, $b$, $s$,
$m^2_Z$, $\Gamma_Z$. Then the cross-section is proportional to
\cite{landau}
\begin{eqnarray}
\sigma\sim |T_0|^2+2\sqrt{s-4m_N^2}\Im[T_0T_1^*],
~~~~~s>4m_N^2\nonumber\\ |T_0|^2-2\sqrt{4m_N^2-s}\Re[T_0T_1^*],
~~~~~s<4m_N^2.
\end{eqnarray}
The form of the cross-section energy behavior near threshold is
defined by the value of the angle, $arg(T_0)-arg(T_1)$ (see Fig. 2
) \cite{landau}. In all cases there are two branches lying on both
sides of common vertical tangent. Thus, the existence of the
reaction threshold leads to the appearance of the characteristic
energy dependence of the cross section. The cross section near
threshold is the linear function of $\sqrt{s-4m_N^2}$ with
different slopes under and above threshold. The existence of the
square root branch point, $s=4m_N^2$, prevents the amplitude
expansion near branch point in Taylor series.

\begin{figure}[h!]
\begin{tabular}{cccc}
\epsfxsize3cm\epsffile{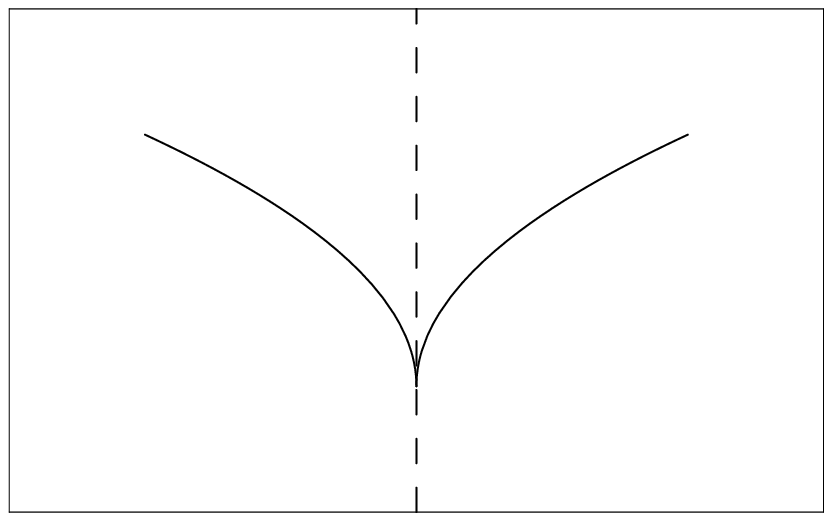}&
\epsfxsize3cm\epsffile{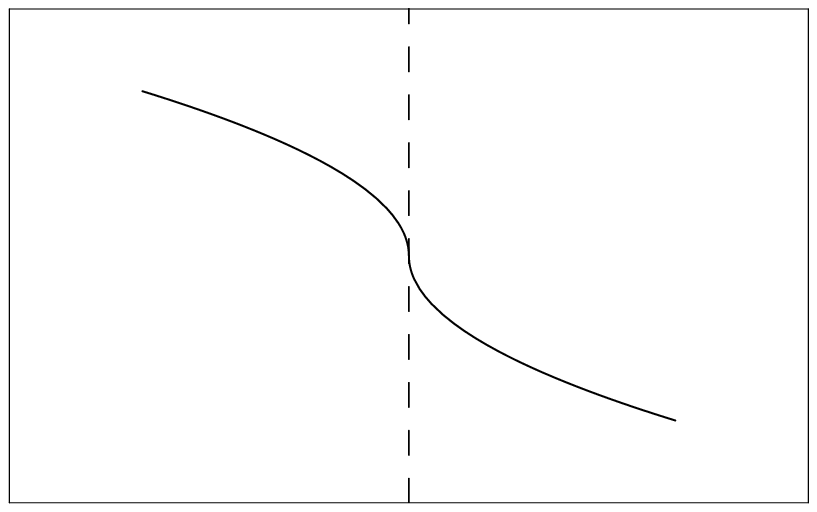}&
\epsfxsize3cm\epsffile{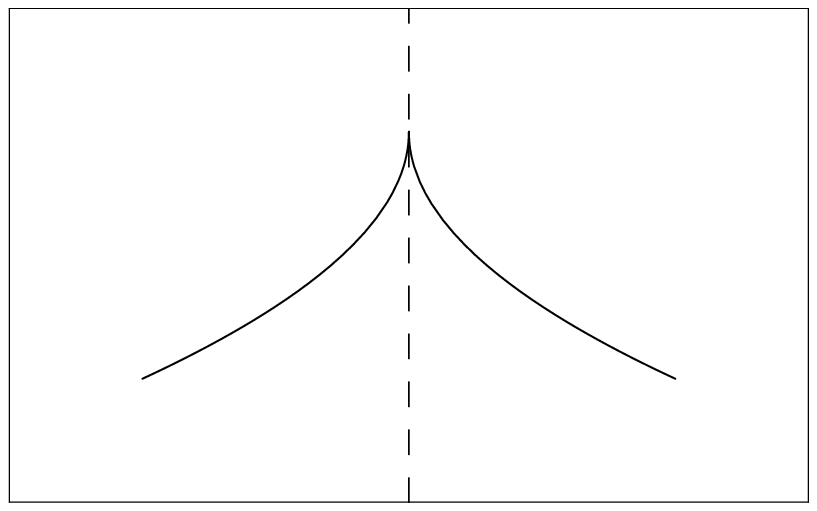}&
\epsfxsize3cm\epsffile{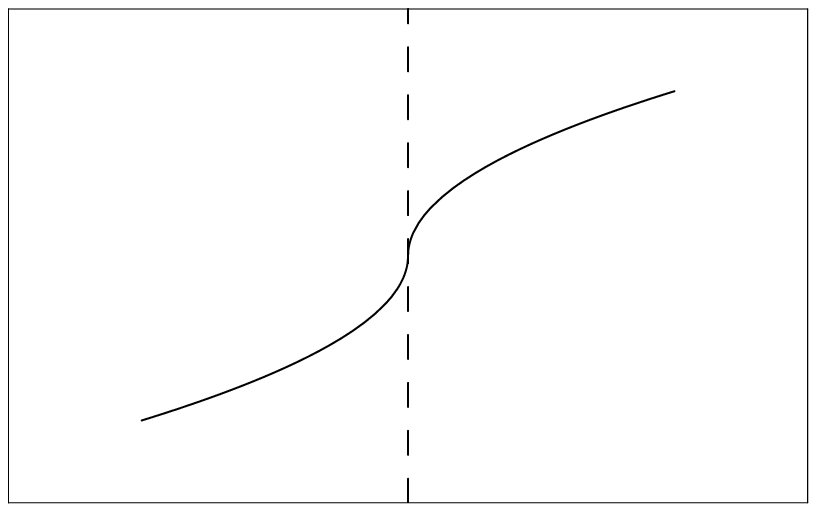}\\
a)&b)&c)&d)
\end{tabular}
\caption{Different cases of cross section behavior near threshold.
Vertical axis is $\sigma$, while horizontal one is $s$; dashed
line crosses horizontal axis at $4m_N^2$}
\end{figure}

Below we will consider the case when $R\equiv Z$.

\section{Polarization operator.}

The cross-section of $e^+e^-\rightarrow Z\rightarrow hadrons$ near
Z-resonance is well described by the Breit-Wigner formula
\cite{10}
\begin{equation}
\sigma_h^{SM}=\frac{12\pi\Gamma_e \Gamma_h} {|p^2-m_Z^2+i\Gamma^{SM}_Z
p^2/m_Z|^2}\frac{p^2}{m_Z^2}, \label{sigma}
\end{equation}
where $p=p_1+p_2$, $p_1$ and $p_2$ are momenta of initial electron
and positron, $m_Z$ is the mass of Z boson, $\Gamma_e$ is the
width of $Z\rightarrow e^+e^-$ decay, $\Gamma_h$ is the width of
$Z\rightarrow hadrons$, $\Gamma^{SM}_Z$ is the total width of Z in the SM.

The 4th generation contributes to the Z-boson polarization
operator. This contribution can be accounted for in the expression
(\ref{sigma}) by replacing the denominator:
\begin{eqnarray}
|p^2-m_Z^2+i\Gamma^{SM}_Z p^2/m_Z|^2\rightarrow \nonumber\\
|p^2-m_Z^2+i\Gamma_Z p^2
/m_Z+\Sigma^{(4th)}_Z(p^2)-\Re[\Sigma^{(4th)}_Z(m_Z^2)]|^2, \label{denom}
\end{eqnarray}
where the subtraction is performed due to the fact that we follow
the approach of \cite{1,2,4} and use for $m_Z$, $\alpha$ and $G_F$
experimental values. Thus the renormalization  scheme is on-shell
one. The real part of the polarization operator at $s=m_Z^2$ is
subtracted in order to avoid the shifting of Z-boson mass. It is
due to the fact that the real part of the polarization operator
contributes to $m_Z$.
\begin{equation}
\Sigma^{(4th)}_Z(p^2)=\Sigma^{(N)}_Z(p^2)+\Sigma^{(E)}_Z(p^2)+
\Sigma^{(U)}_Z(p^2)+\Sigma^{(D)}_Z(p^2)
\end{equation}
is the contribution of the 4th generation to Z polarization
operator. The contribution of the $N\bar{N}$ channel into Z width is taken 
into account by the imaginary part of $\Sigma^{(N)}_Z(p^2)$. 

Note, that $\Gamma_Z$ in (\ref{denom}) includes decays of Z into particles of 
the first three generations. The 4th generation also influences $\Gamma_Z$ in 
non-direct way. It is due to the fact that the polarization operators
of gauge bosons enter the radiative corrections for $g_A$ and
$g_V$, which in their turn enter the amplitude of the Z-boson
decay into fermion-antifermion pair:
\begin{equation}
M(Z\rightarrow f\bar{f})=\frac{1}{2}\bar{f}Z_\alpha\bar{\psi}_f
(\gamma_\alpha g_{Vf}+\gamma_\alpha\gamma_5 g_{Af})\psi_f,
\end{equation}
where $\bar{f}^2=4\sqrt{2}G_\mu m^2_Z=0.54866(4)$, $G_\mu$ is Fermi
coupling constant. In the case of the Z decay into $\nu\bar{\nu}$
the contribution of final state interaction equals zero and
\begin{equation}
\Gamma_\nu=4\Gamma_0(g_{V\nu}^2+g_{A\nu}^2),
\end{equation}
where $\Gamma_0=G_\mu m_Z^3/24\sqrt{2}\pi$ is the so called
"standard" width. If we neglect the masses of neutrinos then
$g_{V\nu}=g_{A\nu}=g_\nu$. For the decay into any pair of charged
leptons we get:
\begin{equation}
\Gamma_l=4\Gamma_0\left[g_{Vl}^2\left(1+\frac{3\bar{\alpha}}{4\pi}\right)
+g_{Al}^2\left(1+\frac{3\bar{\alpha}}{4\pi}
-6\frac{m_l^2}{m_Z^2}\right)\right],
\end{equation}
where $m_l$ is the mass of the lepton, $\bar{\alpha}\equiv
\alpha(m_Z^2)=[128.896(90)]^{-1}$. The situation slightly changes in
the case of $Z\rightarrow q\bar{q}$. There appear the radiative
corrections ($R_{Vq}$ and $R_{Aq}$) due to gluon exchange and
emission in the final state:
\begin{equation}
\Gamma_q=12\Gamma_0\left(g_{Vq}^2R_{Vq}+g_{Aq}^2R_{Aq}\right).
\end{equation}

According to the results of \cite{8} the one-loop expressions for
$g_{Al}$ and $R_l=g_{Vl}/g_{Al}$ are
\begin{equation}
g_{Al}=-\frac{1}{2}-\frac{3\bar{\alpha}}{64\pi s^2c^2}V_A, ~~~
R_l=1-4s^2+\frac{3\bar{\alpha}}{4\pi(c^2-s^2)}V_R,
\end{equation}
\begin{equation}
g_\nu=\frac{1}{2}+\frac{3\bar{\alpha}}{64\pi s^2c^2}V_\nu,
\end{equation}
and in the case of quarks
\begin{equation}
g_{Aq}=T_{3q}\left[1+\frac{3\bar{\alpha}}{32\pi
s^2c^2}V_{Aq}\right],~~~
R_q=1-4|Q_q|s^2+\frac{3\bar{\alpha}|Q_q|}{4\pi(c^2-s^2)}V_{Rq},
\end{equation}
where $c\equiv \cos\theta_{eff}$ and $s\equiv \sin\theta_{eff}$.

The exact expressions for $V_A$, $V_R$, $V_{Aq}$ and $V_{Rq}$ in
SM can be found in \cite{8}.

The 4th generation particles contribute to physical observables
through polarization operators of gauge bosons, as it was
mentioned above. This gives corrections $\delta V_i$ to the
functions $V_i$ ($i=A,~R$) \cite{1}.
\begin{eqnarray}
\frac{3\bar{\alpha}}{16\pi s^2c^2}\delta V_A=\Pi_Z^{(4th)}(m^2_Z)
-\Pi^{(4th)}_W(0) -\Sigma^{(4th)\prime }_Z(m^2_Z), \nonumber\\
\frac{3\bar{\alpha}}{16\pi s^2c^2}\delta
V_R=\left[\Pi_Z^{(4th)}(m^2_Z) -\Pi_W^{(4th)}(0)
-\Sigma^{(4th)\prime }_\gamma(0)\right] \nonumber\\
-\frac{sc(c^2-s^2)}{s^2c^2}\Pi_{\gamma Z}^{(4th)}(m^2_Z), \label{dv}
\end{eqnarray}
where $\Pi_Z(m_Z^2)=\Sigma_Z(m^2_Z)/m_Z^2$. Note that all singular
terms, proportional to $1/\varepsilon$ ($\varepsilon=D-4$), in
right hand side of eq. (\ref{dv}), arising from polarization
operators, cancel, as it was shown in \cite{8}. Thus, the
expressions (\ref{dv}) are finite. However, these formulae work
well for the new particles being much heavier than Z-boson. If
$m_N\rightarrow m_Z/2$ then $\Sigma^{(4th)\prime}_Z(m^2_Z)$ tends
to infinity. $\Sigma^{(4th)\prime}_Z(m^2_Z)$ comes from the Taylor
expansion of Z-boson polarization operator near $m_Z$. In order to
get rid of this unphysical infinity, which arise due to the fact
that the expansion is performed at the branch point of the
polarization operator, we use the exact expression for the
contribution of 4th generation particles to the Z-boson
polarization operator.

For the contribution of the 4th generation particles we get
\begin{eqnarray}
\Pi^{(\phi)}_Z(p^2)=\frac{N_c\bar{\alpha}}{8\pi s^2
c^2}\left[\Delta_\phi+4g_{A\phi}^2
\frac{m_\phi^2}{p^2}F\left(\frac{m_\phi^2}{p^2}\right)\right. \nonumber\\
\left.-\frac{2(g_{A\phi}^2+g_{V\phi}^2)}{3}\left((1+2\frac{m_\phi^2}{p^2})
F\left(\frac{m_\phi^2}{p^2}\right)+\frac{1}{3}\right)\right]
\label{pi_N},
\end{eqnarray}
where $\phi=$N, E, U, D; $N_c=3$ for quarks and $N_c=1$ for
leptons; $\Sigma_Z^{(\phi)}(p^2)=p^2\Pi_Z^{(\phi)}(p^2)$;
\begin{equation}
F(x)=\left[-2+2\sqrt{4x-1}\arctan{\frac{1}{\sqrt{4x-1}}}\right],
\end{equation}
$\Delta_\phi$ are singular parts:
\begin{equation}
\Delta_\phi=2\left(\frac{1}{3}(g_{A\phi}^2+g_{V\phi}^2)-2g_{A\phi}^2\frac{m_\phi^2}{p^2}\right)
\left(\frac{1}{\varepsilon}-\gamma+\ln
4\pi-\ln\frac{m_\phi^2}{\mu^2}\right),
\end{equation}
where $\varepsilon\rightarrow 0$,
$\gamma=-\Gamma^\prime(1)=0.577...$, $\mu$ is the parameter with
dimension of mass, which is needed to preserve the dimensionality
of the initial integral.

Let's consider the contribution of N to $g_A$ and $g_V$.
$\Sigma_Z^\prime(m_Z^2)$ arises in $g_A$ due to the
renormalization of Z-boson wave function. In order to avoid infinities
we will expand near $m_Z$ only singular part of the polarization
operator, i. e.

$$ p^2-m^2_{Z}+\Sigma_Z(p^2)=  $$
\begin{equation}
=(1+\left.\Sigma_{Z}^\prime(m_Z^2)\right|_s)[p^2-m^2(1-\left.\Pi_{Z}(m^2)\right|_s
-\frac{p^2}{m^2}\left.\Pi_{Z}(p^2)\right|_{FP})], \nonumber
\end{equation}
where index $s$ denotes singular part and index $FP$ -- finite
part. Then in equations (\ref{dv}) we should replace
$\Sigma^\prime(m_Z^2)_Z$ by
$\left.\Sigma^\prime(m_Z^2)_Z\right|_s$ and $\Pi(m_Z^2)_Z$ by
$\left.\Pi_Z(m^2)_Z\right|_s
+\frac{p^2}{m_Z^2}\left.\Pi_{Z}(p^2)\right|_{FP}$. The combination
of singular terms in the resulting expression is the same as in
(\ref{dv}) and that's why they cancel each other.

Let's consider the expression for the $e^+e^-\rightarrow hadrons$ cross
section in the presence of the 4th generation. The singular part of the
polarization operator is absorbed in partial widths $\Gamma_e$ and $\Gamma_h$
and also in total width
of Z, $\Gamma_Z$.
\begin{eqnarray}
\frac{\Gamma_e^0\Gamma_h^0}{|p^2-m_Z^2+i\Gamma_Z^0
p^2/m_Z+\Sigma^{(4th)}(p^2)-\Re[\Sigma^{(4th)}(m_Z^2)]|^2}
\nonumber\\
=\frac{\Gamma_e^0\Gamma_h^0}{(1+\left.\Sigma^{\prime}\right|_s)^2|p^2-m_Z^2+\frac{i\Gamma_Z^0
p^2}{(1+\left.\Sigma^{\prime}\right|_s)m_Z}+(\Sigma^{(4th)}(p^2)-\Re[\Sigma^{(4th)}(m_Z^2)])_{FP}
|^2} \nonumber\\ =\frac{\Gamma_e\Gamma_h}{|p^2-m_Z^2+i\Gamma_Z
p^2/m_Z+(\Sigma^{(4th)}(p^2)-\Re[\Sigma^{(4th)}(m_Z^2)])_{FP}|^2},
\end{eqnarray}
where $\Gamma_e=\Gamma_e^0/(1+\left.\Sigma^{\prime}\right|_s)$,
$\Gamma_h=\Gamma_h^0/(1+\left.\Sigma^{\prime}\right|_s)$,
$\Gamma_Z=\Gamma_Z^0/(1+\left.\Sigma^{\prime}\right|_s)$ and we
used the decomposition
$$\Sigma^{(4th)}(p^2)-\Re[\Sigma^{(4th)}(m_Z^2)]$$ $$
=(\Sigma^{(4th)}(p^2)-\Re[\Sigma^{(4th)}(m_Z^2)])_s
+(\Sigma^{(4th)}(p^2)-\Re[\Sigma^{(4th)}(m_Z^2)])_{FP}$$
$$=\left.\Sigma^\prime(m_Z^2)\right|_s(p^2-m_Z^2)
+(\Sigma^{(4th)}(p^2)-\Re[\Sigma^{(4th)}(m_Z^2)])_{FP}$$ The
cancellation of the singularities is due to the fact that
$\Gamma_e$, $\Gamma_h$ and  $\Gamma_Z$ are proportional to
$\bar{f}^2_0$, which can be rewritten in terms of $G_\mu$, $m_Z$
and polarization operators as in \cite{8}:
\begin{equation}
\bar{f}^2_0=4\sqrt{2}G_\mu m_Z^2[1-\Pi_W(0)+\Pi_Z(m^2_Z)-D],
\end{equation}
where $D$ comes from the radiative corrections to $G_\mu$
\cite{8}. If we divide $\bar{f}^2_0$ by $(1+\Sigma^{(s)\prime})$
then the resulting expression will be finite, due to the fact that
all singular terms cancel. Thus, in the expression for the cross
section of $e^+ e^-\rightarrow hadrons$ there are no singular
terms.

We should note, that there is an ambiguity in the definition of
singular and finite parts of the polarization operator. The
constant term, proportional to $\bar{\alpha}$, can be added to
singular part and subtracted from finite part: $$
\Pi_Z(p^2)=\left(\left.\Pi_Z(p^2)\right|_s+a\bar{\alpha}\right)+
\left(\left.\Pi_Z(p^2)\right|_{FP}-a\bar{\alpha}\right). $$
However, this ambiguity does not affect the expression for the
cross section. It is obvious from the following expressions:
\begin{eqnarray}
\frac{\Gamma_e^0\Gamma_h^0}{(1+\left.\Sigma^{\prime}\right|_s+a\bar{\alpha})^2|p^2-m_Z^2+\frac{i\Gamma_Z^0
p^2}{(1+\left.\Sigma^{\prime}\right|_s)m_Z}+\hat{\Sigma}-a\bar{\alpha}(p^2-m_Z^2)
|^2} \nonumber\\
=\frac{\Gamma_e^{\prime}\Gamma_h^{\prime}}{|(p^2-m_Z^2)(1-a\bar{\alpha})+\frac{i\Gamma_Z^{\prime}
p^2}{m_Z}+\hat{\Sigma}|^2}
=\frac{\Gamma_e^{\prime}\Gamma_h^{\prime}/(1-a\bar{\alpha})^2}{|(p^2-m_Z^2)+\frac{i\Gamma_Z^{\prime}
p^2}{(1-a\bar{\alpha})m_Z}+\hat{\Sigma}|^2} \nonumber
\end{eqnarray}
$$ =\frac{\Gamma_e\Gamma_h}{|(p^2-m_Z^2)+\frac{i\Gamma_Z
p^2}{m_Z}+\hat{\Sigma}|^2}, \nonumber $$ where
$\hat{\Sigma}=(\Sigma^{(4th)}(p^2)-\Re[\Sigma^{(4th)}(m_Z^2)])_{FP}$,
$$ \Gamma_i=\frac{\Gamma_i^{\prime}}{1-a\bar{\alpha}}
=\frac{\Gamma_i^0}{(1+\left.\Sigma^{\prime}\right|_s+a\bar{\alpha})(1-a\bar{\alpha})}=
\frac{\Gamma_i^0}{1+\left.\Sigma^{\prime}\right|_s+a\bar{\alpha}-a\bar{\alpha}}
$$ $$ =\frac{\Gamma_i^0}{(1+\left.\Sigma^{\prime}\right|_s)}, $$
where $i=e,h,Z$.

\section{Z-lineshape in the presence of 4th generation.}

According to the results of the previous section, the amplitude of
$e^+e^-\rightarrow hadrons$ is proportional to
\begin{equation}
A_h\sim\left[p^2-m_Z^2+i\Gamma_Z^0
p^2/m_Z+\Sigma^{(4th)}_Z(p^2)-\Re(\Sigma^{(4th)}_Z(m_Z^2))\right]^{-1}.
\label{amp}
\end{equation}
In order to study the behavior of the cross section near the threshold
qualitively, we shall neglect the contributions of U, D and E to
$\Sigma^{(4th)}_Z(p^2)$ as well as the contribution of all 4th generation
particles to $g_A$ and $g_V$. Then the expression (\ref{amp}) takes the
following form:
\begin{equation}
A_h\sim\left[p^2-m_Z^2+i\Gamma_Z
p^2/m_Z+(\Sigma^{(N)}_Z(p^2)-\Re[\Sigma^{(N)}_Z(m_Z^2)])_{FP}\right]^{-1}.
\end{equation}
Expanding this expression near the threshold of $N\bar{N}$
production ($p^2=4m_N^2$, $A_h\sim T_0+iT_1\sqrt{p^2-4m^2_N}$) we obtain for
the cross section the following behaviour
\begin{equation}
\sigma_{p^2<4m_N^2}\sim\frac{1}{\gamma}
\left(1+\frac{\bar{f}^2 m_Z^2}{64\pi}
\frac{(4m_N^2-m_Z^2)\sqrt{\frac{4m_N^2}{p^2}-1}}{\gamma}\right),
\label{s1}
\end{equation}
\begin{equation}
\sigma_{p^2>4m_N^2}\sim\frac{1}{\gamma}
\left(1-\frac{\bar{f}^2 m_Z^2}{64\pi}
\frac{m_Z\Gamma_Z\sqrt{1-\frac{4m_N^2}{p^2}}}{\gamma}\right),
\label{s2}
\end{equation}
where $\gamma=(4m_N^2-m_Z^2)^2+(m_Z\Gamma_Z)^2$ and the second
terms in the brackets in eqns. (\ref{s1}, \ref{s2}) are
proportional to $\Re[T_0T_1^*]$ and $\Im[T_0T_1^*]$ respectively.
As it was mentioned in section 2 the form of the $p^2$ dependence
of the cross section near the threshold is determined by the
relative phase of $T_0$ and $T_1$. In our case we have two types
of casps (see Figs. 3), which correspond to $arg T_0-arg T_1$
lying in the third quadrant for $4m^2_N<m_Z^2$ and in fourth
quadrant for $4m^2_N>m_Z^2$, or to $\Re[T_0T_1^*]$ and
$\Im[T_0T_1^*]$ being negative for $4m^2_N<m_Z^2$ and
$\Re[T_0T_1^*]$ being positive, while $\Im[T_0T_1^*]$ being
negative for $4m^2_N>m_Z^2$. It can also be seen from Figs. 3 that
the cross section of $e^+e^-\rightarrow hadrons$ decreases above
the threshold in accordance with the unitarity.

\begin{figure}[h!]
\begin{tabular}{ccc}
\epsfxsize6cm\epsffile{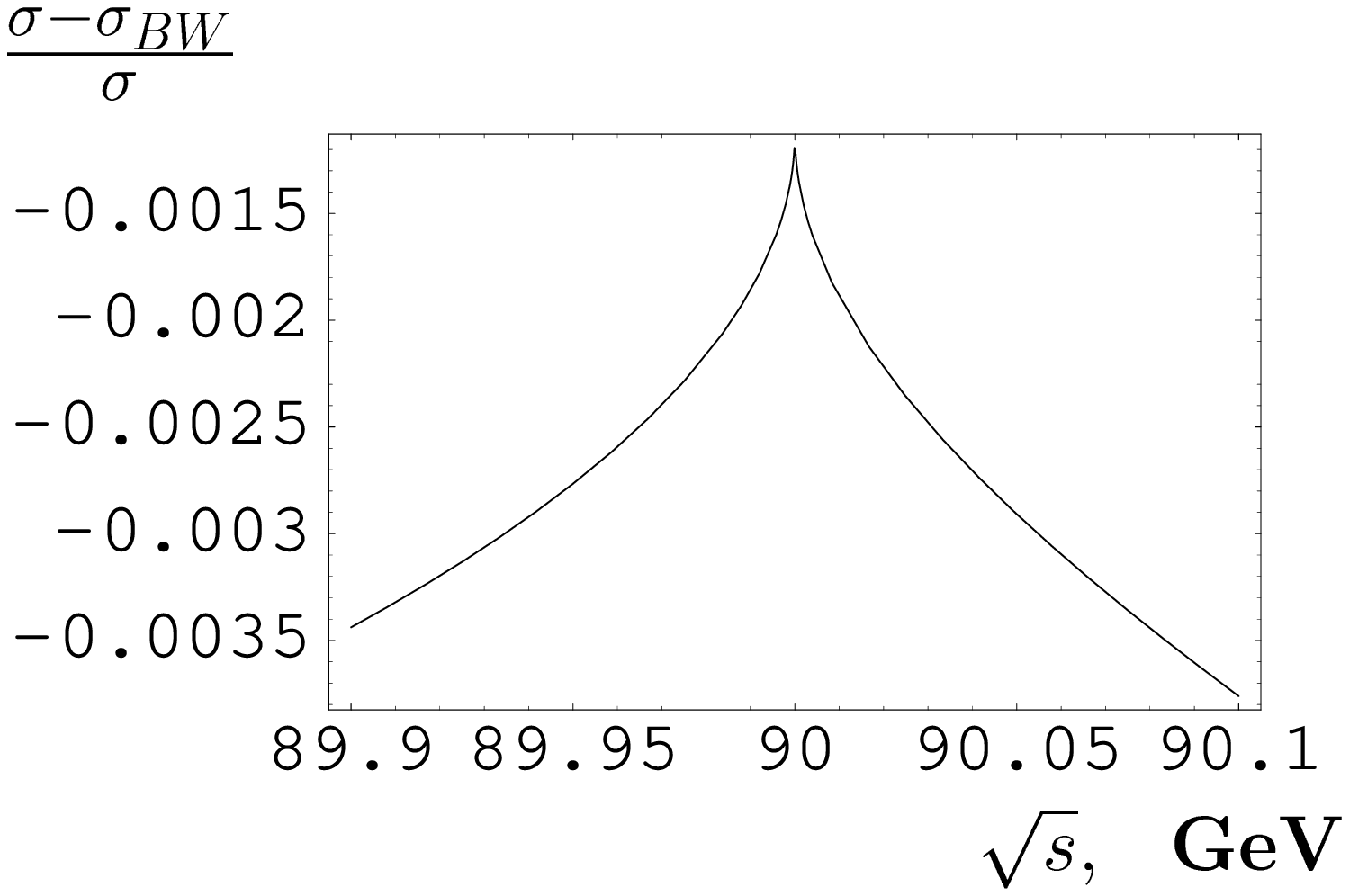}&
\epsfxsize6cm\epsffile{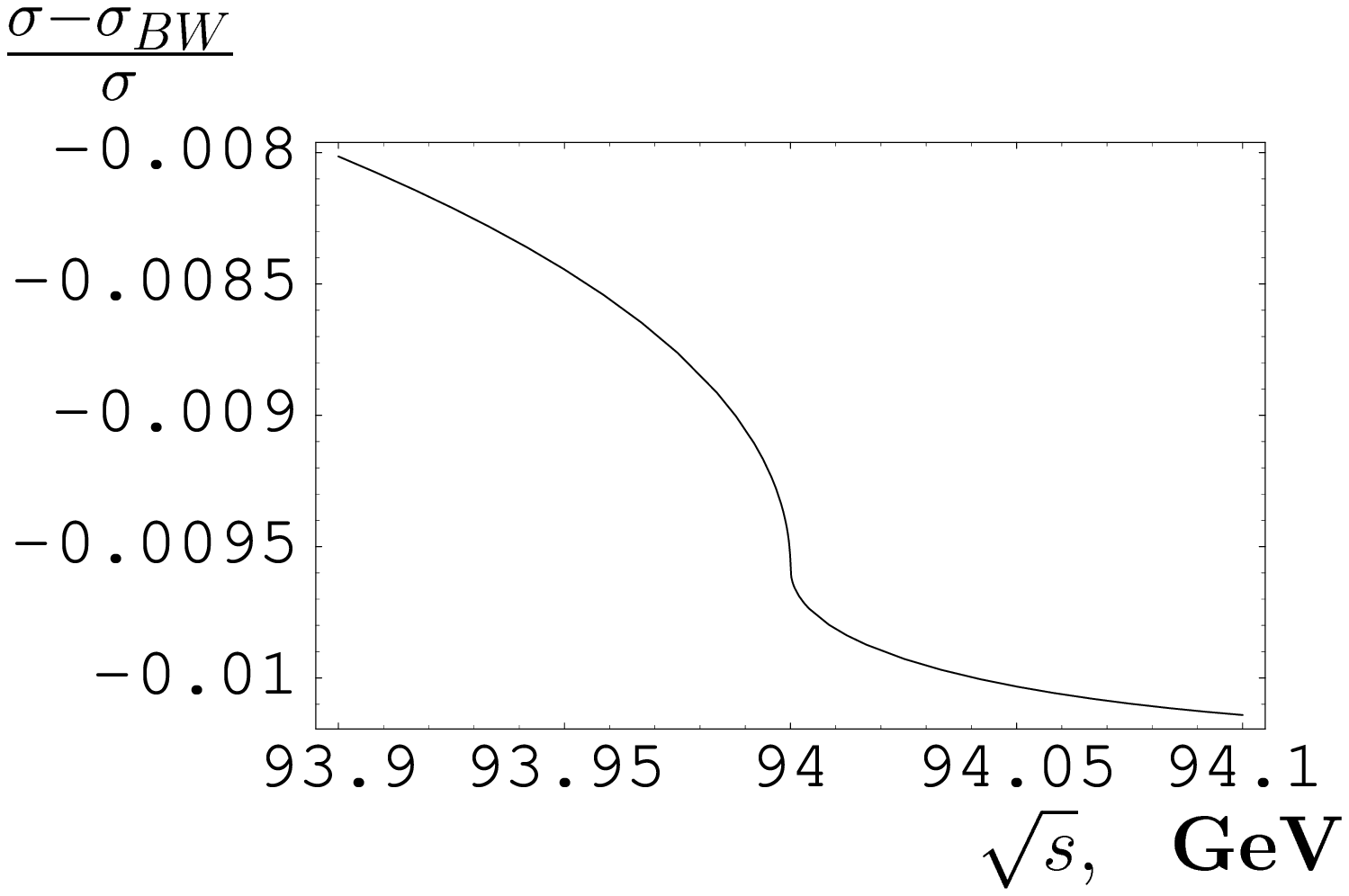}\\ a)&b)
\end{tabular}
\caption{The dependence of relative departure of the
$e^+e^-\rightarrow hadrons$ cross section in the presence of 4th
generation from the SM prediction on the c.m.  energy of $e^+e^-$
for $m_N=45$ GeV (a) and $m_N=47$ GeV (b).}
\end{figure}

Though the change of Z-lineshape due to these casps is very small
compared to the pure Breit-Wigner curve, as it is shown in Fig. 3,
this effect may manifest itself when comparing the theoretical
predictions with the experimental data. It is due to the fact,
that Z-lineshape is measured with very high precision. In the next
section we will compare the theoretical cross section with the
experimental data.

\section{Comparison with the experiment.}

The experimental data on the cross section of $e^+e^-\rightarrow
hadrons$ reaction is usually presented in the form, that includes
the electromagnetic corrections, i. e. initial and final state
interactions and photon emission \cite{9}. In order to compare our
formulae for the cross section with experimental ones, we use the
ZFITTER code \cite{10}, which takes into account these corrections. We use the 
following inputs:
$$
m_Z=91.1882(22)~\mbox{GeV},~~m_t=175(4.4)~\mbox{GeV},~~\bar{\alpha}=1/128.918(45),
$$
$$
\alpha_s=0.1182(27),~~m_H=120~\mbox{GeV}.
$$
With the reasonable assumption that the inital and final
state radiation effects are not significantly modified by fourth
generation we can calculate the cross section
\begin{equation}
\sigma^{th}_h=\sigma_h\frac{\sigma_h^{ZF}}{\sigma_h^{SM}},
\end{equation}
where $\sigma_h^{ZF}$ is the result of ZFITTER code and
\begin{equation}
\sigma_h=\frac{12\pi\Gamma_e\Gamma_h}{|p^2-m_Z^2+i\Gamma_Z
p^2/m_Z+(\Sigma^{(4th)}(p^2)-\Re[\Sigma^{(4th)}(m_Z^2)])_{FP}|^2}\frac{p^2}{m_Z^2},
\end{equation}
at values of c. m. energy, at which the experimental values of
cross section were measured \cite{9}. There are 35 experimental
points from 1995 data, which we use. These points are extracted
from Fig. 2 of \cite{9} and presented in Table 1. We took only the
points corresponding to 1993-1995 set, due to the fact that they
are measured with higher precision than the 1991-1993 set. Then we
calculate the $\chi^2/n_{d.o.f.}$, where $n_{d.o.f.}=35-N$, $N$ is
the number of fitted parameters\footnote{In our case $N=1$,
because only heavy neutrino mass is a free parameter, all other parameters
are fixed.}, and
\begin{equation}
\chi^2=\sum_{i=1}^{35}\left(\frac{\sigma_h^{th}-\sigma_h^{exp}}
{\delta\sigma_h^{exp}}\right)^2,
\end{equation}
$\sigma_h^{exp}$ is the experimental value of cross section and
$\delta\sigma_h^{exp}$ is its error, in order to determine at what
confidence level the 4th generation is excluded by the
experimental data. However, the bound on N mass from below depends
on the higgs mass and mass splittings between U and D quarks and
between E and N leptons. The effects of varying $m_H$,
$|m_U-m_D|$, and $|m_E-m_N|$ are not independent. As it was shown
in \cite{5}, the increase of $|m_U-m_D|$ or $|m_E-m_N|$ can be
compensated by the increase of higgs mass. This leads to the
appearance of $\chi^2_{min}$ valleys. It can be seen from Figs.
4a) and 4b), where the dependence of $\chi^2$ on ${m_H,
|m_U-m_D|}$ (a) and ${m_H, |m_E-m_N|}$ (b) is shown for $m_N=49$
GeV.

\begin{figure}[h!]
\begin{tabular}{ccc}
\epsfxsize6cm\epsffile{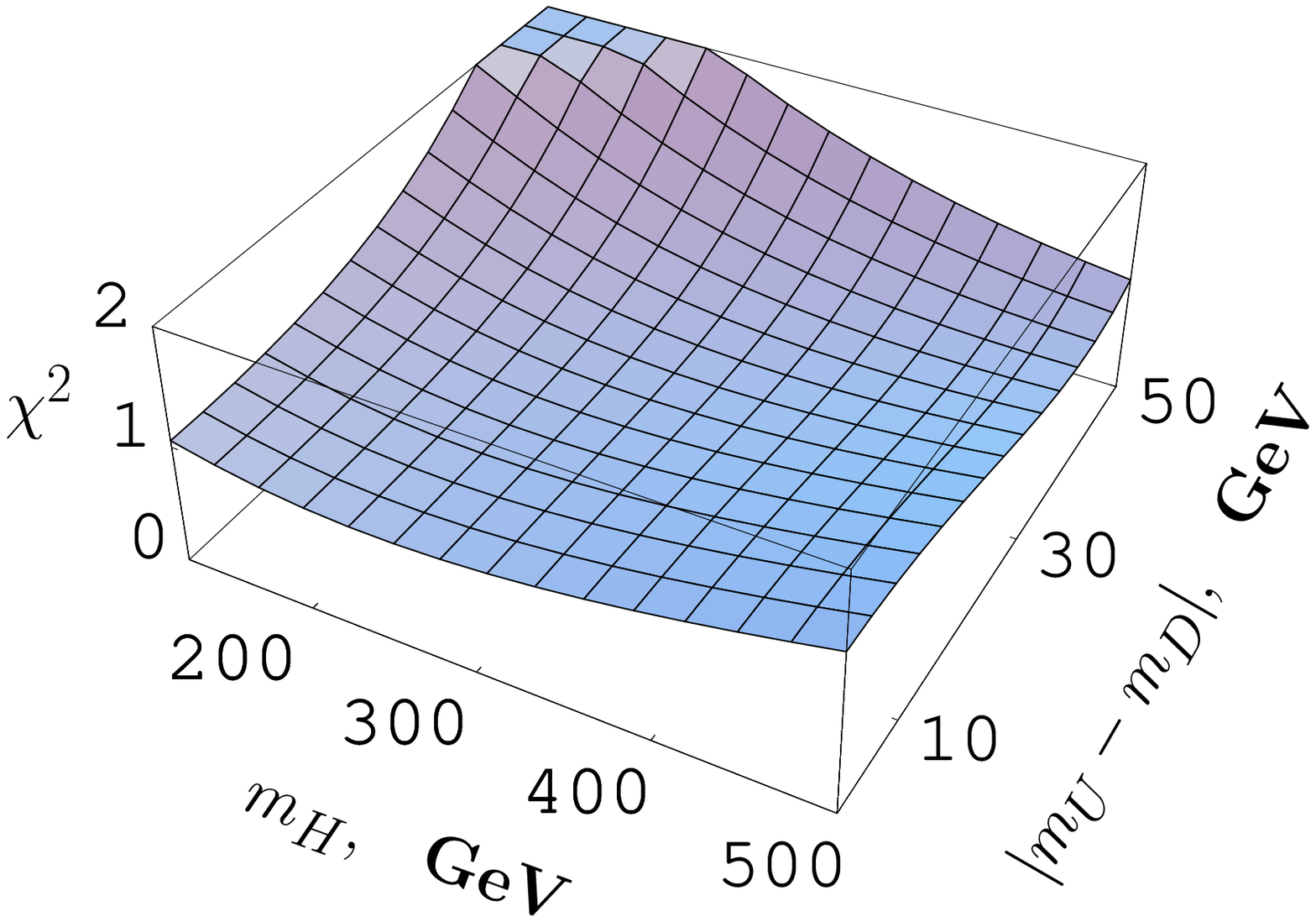}&
\epsfxsize6cm\epsffile{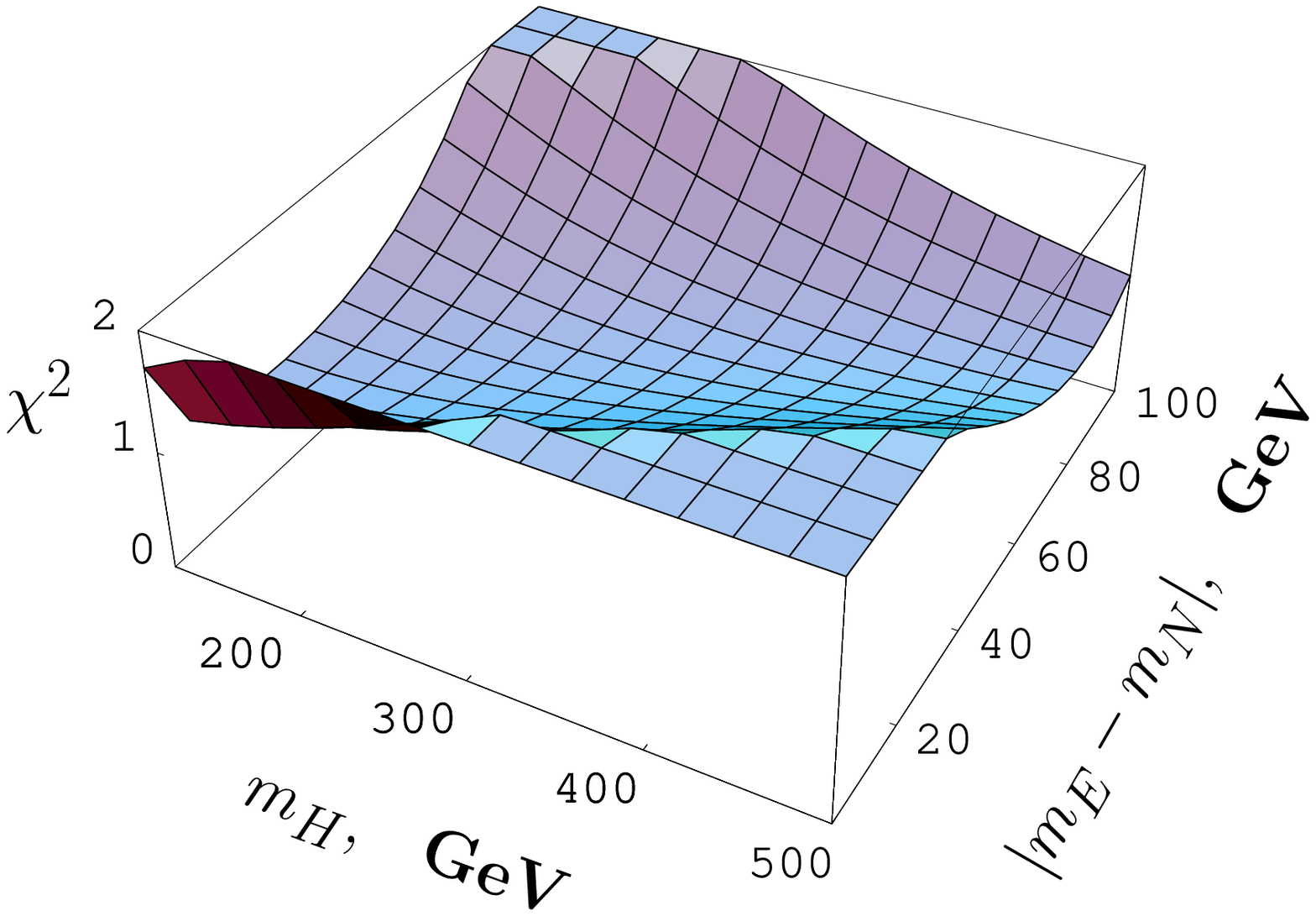}\\ a)&b)
\end{tabular}
\caption{The dependence of $\chi^2$ on $m_H$, $|m_U-m_D|$ (a) and
on $m_H$, $|m_E-m_N|$.}
\end{figure}

If we use then the LEP II bound $m_E>100$ GeV and the results of
\cite{5}, that the best fit of electroweak data corresponds to the
light E near the bound and $m_N\approx 50$ GeV, then we have only
two parameters, $m_H$ and $|m_U-m_D|$ that affect the bound on
$m_N$.  As it can be seen from Fig. 5a) the best fit is acquired
for $0.11m_H -19.7<|m_U-m_D|<0.12m_H-9.2$. In this region of
masses we calculate $\chi^2$ and find that the 4th generation is
excluded at 95\% C. L. for $m_N<46.7\pm0.2$ GeV. The theoretical
uncertainty is caused by the varying of $|m_U-m_D|$ from 0 to 50
GeV, as well as by the uncertainties of the input parameters of
ZFITTER, which were also used when calculating $\sigma_h$ and
$\sigma_h^{SM}$. The main contribution to the theoretical
uncertainty comes from $m_t$ and $\alpha_S$. The variation from 0
to 50 GeV is chosen, because the quality of the fit is fast
worsening for $|m_U-m_D|>50$ GeV.

\begin{figure}[h!]
\begin{tabular}{ccc}
\epsfxsize6cm\epsffile{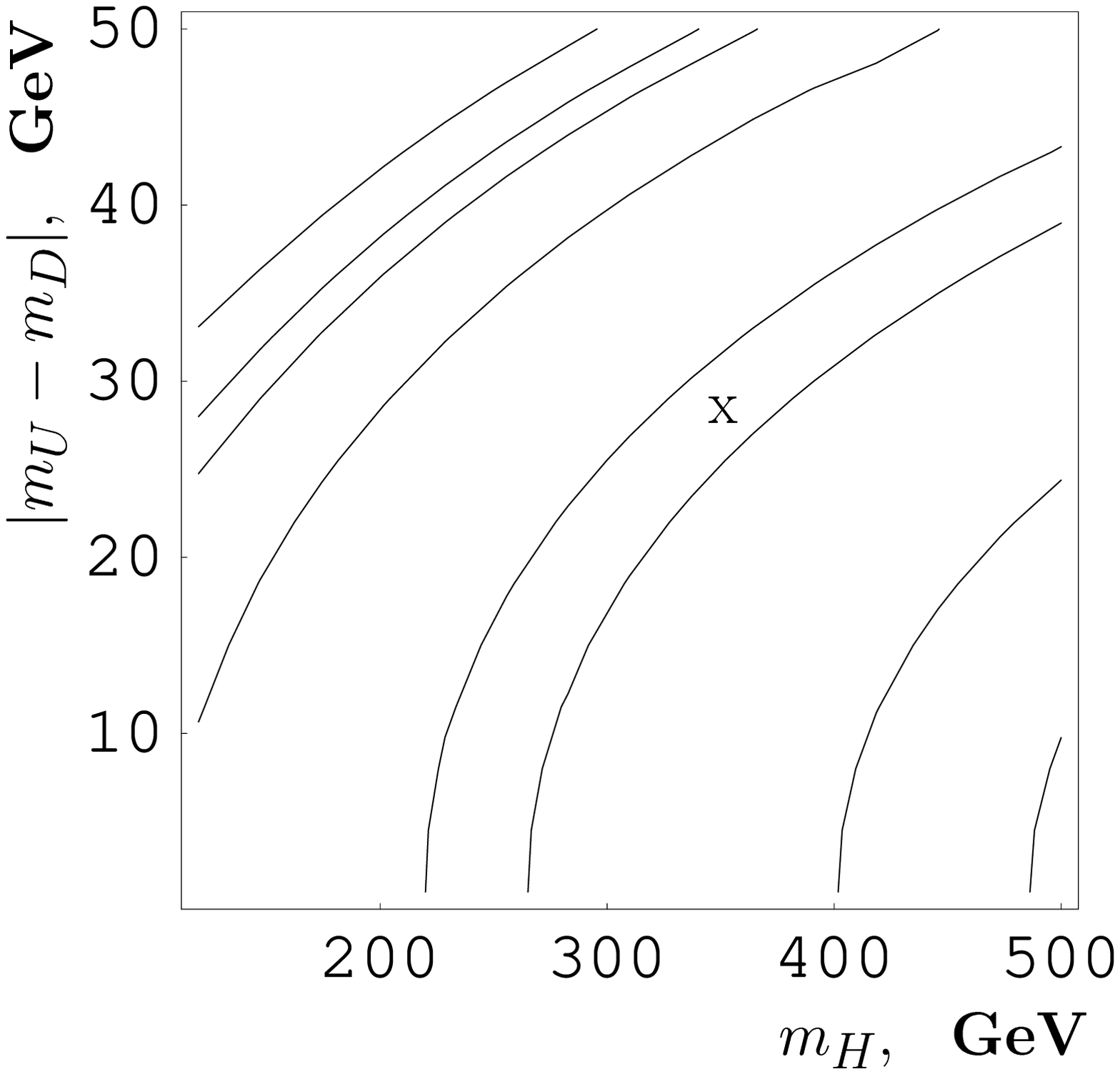}&
\epsfxsize6cm\epsffile{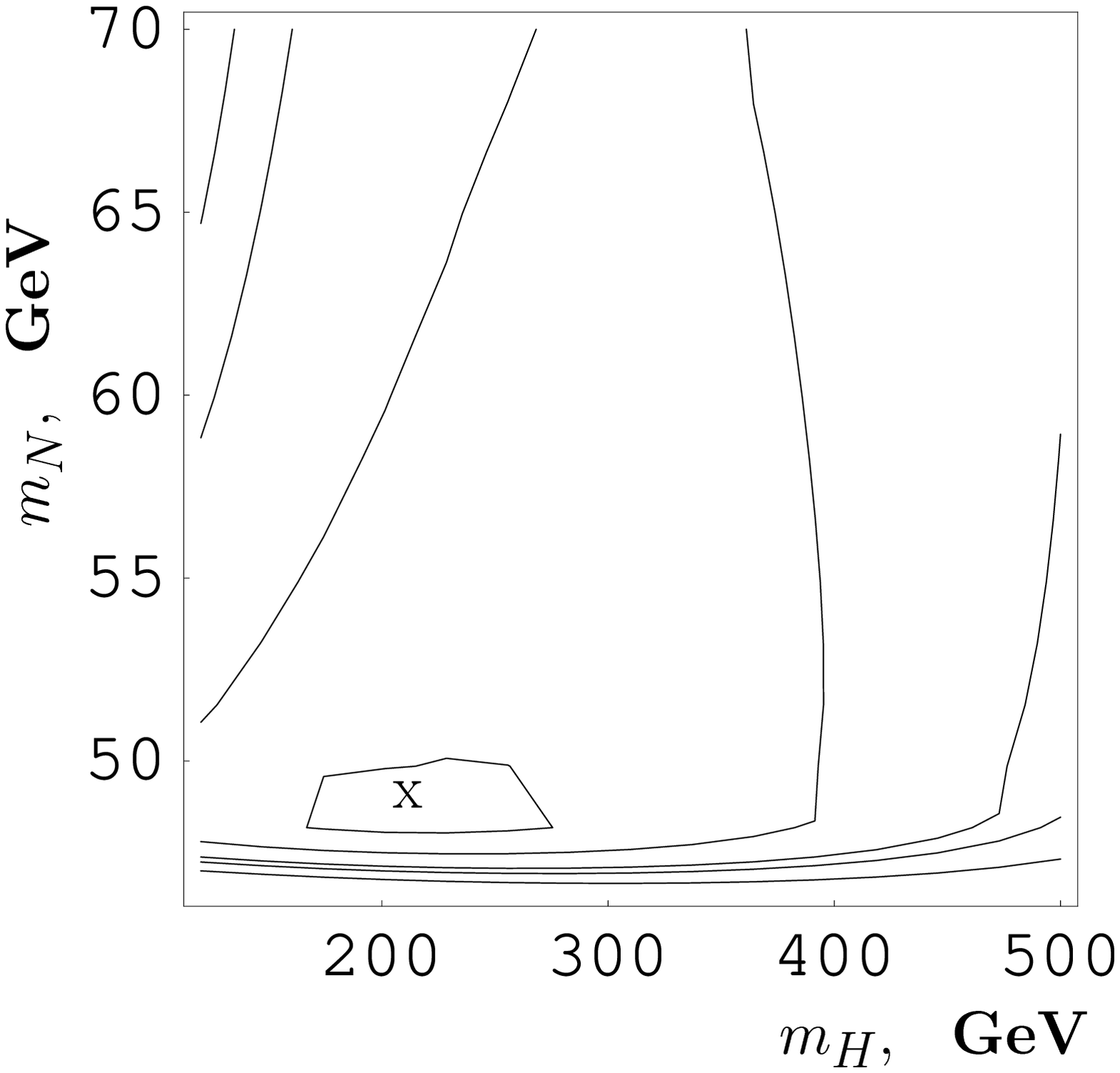}\\ a)&b)
\end{tabular}
\caption{a) Exclusion plot on the plane $m_H$, $|m_U-m_D|$ for
$m_N=49$ GeV; $\chi^2_{min}=0.85$ denoted by cross b) Exclusion
plot on the plane $m_H$, $m_N$ for $|m_U-m_D|=10$ GeV;
$\chi^2_{min}=0.85$ denoted by cross. Solid lines represents the
borders of $1 \sigma$, $2 \sigma$, $3 \sigma$, $4 \sigma$ and $5
\sigma$ regions.}
\end{figure}

In order to illustrate the dependence of the fit quality on the
higgs mass we study $\chi^2(m_N,m_H)$ (see Fig. 5b)). From this
figure it is seen that the 95\% C. L. bound lies below 50 GeV and
slightly varies with the increase of the higgs mass near $m_N=47$
GeV. In this figure we take $|m_U-m_D|=10$ GeV. It can be seen
from Figs. 5 a) and b) that for certain region of 4th generation
particles and higgs masses the quality of the fit can be even
better than in SM. According to the results of \cite{9} the
$\chi^2/n_{d.o.f.}(SM)=1.09$, which corresponds to $2\sigma$
level, while in the presence of 4th generation it is
$\chi^2_{min}/n_{d.o.f.}=0.88$, which is inside $1\sigma$ region.

We should note, that the direct search of heavy neutrinos in
$e^+e^-$ annihilation into a pair of heavy neutrinos with the
emission of the initial state bremsstrahlung photon
($e^+e^-\rightarrow \gamma+Nothing$) could result in the bound $m_N \geq 50$
GeV \cite{2,4} if all four LEP experiments will make a combined analysis 
\cite{6}.

Though this bound on $m_N$ would be slightly better than the one,
obtained in the present paper, the data and the procedure used to
extract the bounds are completely different and independent. 

\section{Conclusions.}

In the present paper we analyzed the dependence of Z-lineshape on the location
of the threshold of $N\bar{N}$ production. We studied the behavior of
$e^+e^-\rightarrow hadrons$ cross section near the threshold of $N\bar{N}$
production and determined how this threshold changes the Z-lineshape. In order
to find the bound on $m_N$, we compared the theoretical
predictions for the Z-lineshape with the experimental data, using
the exact formulae for Z polarization operator, instead of
expanding it into a Taylor series near $m_Z$.

We found that the bound on N mass depends on the higgs mass and
the splittings of 4th generation  quark and lepton masses
($|m_U-m_D|$ and $|m_E-m_N|$). However, the effects caused by them
are not independent, because the increase of mass splittings can
be compensated by the increase of the higgs mass, as it was shown
in \cite{5}. Using the results of \cite{5} we fixed $m_E=100$ GeV.
Then we used the fact that $|m_U-m_D|$ and $m_H$ are not
independent. Thus, we had one free parameter left: $|m_U-m_D|$. We
varied $|m_U-m_D|$ from 0 to 50 GeV and found that the 4th
generation is excluded by the experimental data at 95\% C. L. for
$m_N<46.7\pm0.2$ GeV. The theoretical uncertainty is caused by
the varying of $|m_U-m_D|$ , as well as by the uncertainties of
the input parameters of ZFITTER, which were also used when
calculating $\sigma_h$ and $\sigma_h^{SM}$.

\section*{Acknowledgements}
The authors are grateful to M. Yu. Khlopov and N. Mankoc Borstnik for 
correspondence. This work was partially supported by RFBR (grant N 
00-15-96562).

\newpage
\begin{center}
Table 1. \\ The experimental values of the $e^+e^-\rightarrow
hadrons$ cross section, obtained by ALEPH, DELPHI, L3 and OPAL
collaborations, extracted from Fig. 2 of \cite{9}. $\sqrt{s}$ is
presented in GeV, $\sigma_h$ in nanobarns. The 1993-1995 data set.
\end{center}

\begin{center}
Table 1.1 ALEPH
\end{center}

\begin{tabular}{|c|c|c|c|c|c|c|}
\hline
$\sqrt{s}$& 89.4316& 89.4400& 91.1860& 91.1980& 91.2200& 91.2840 \\ \hline
$\sigma_h$& 9.891& 9.980& 30.500& 30.43& 30.458& 30.555 \\ \hline
$\delta\sigma_h$& 0.043& 0.044& 0.078& 0.032& 0.067& 0.13\\ \hline
\end{tabular}

\begin{tabular}{|c|c|c|c|c|}
\hline
$\sqrt{s}$& 91.2950& 91.3030& 92.9685& 93.0140\\ \hline
$\sigma_h$ & 30.678& 30.660& 14.300& 14.04\\ \hline
$\delta\sigma_h$& 0.078& 0.090& 0.060& 0.056\\ \hline
\end{tabular}

\bigskip

\begin{center}
Table 1.2 DELPHI
\end{center}

\begin{tabular}{|c|c|c|c|c|c|c|}
\hline
$\sqrt{s}$&  89.4307& 89.4378& 91.186& 91.2& 91.203& 91.28\\ \hline
$\sigma_h$& 9.87& 9.93& 30.392& 30.50& 30.46& 30.65\\ \hline
$\delta\sigma_h$& 0.044& 0.056& 0.065& 0.044& 0.19& 0.13\\ \hline
\end{tabular}

\begin{tabular}{|c|c|c|c|c|}
\hline
$\sqrt{s}$ & 91.292& 91.304& 92.966& 93.014\\ \hline
$\sigma_h$  & 30.67& 30.46& 14.35& 13.89 \\ \hline
$\delta\sigma_h$& 0.098& 0.086& 0.044& 0.045 \\ \hline
\end{tabular}

\bigskip

\begin{center}
Table 1.3 L3
\end{center}

\begin{tabular}{|c|c|c|c|c|c|c|}
\hline
$\sqrt{s}$& 89.4497& 89.4515& 91.206& 91.222& 91.297& 91.309
\\ \hline
$\sigma_h$& 10.088& 10.08& 30.358& 30.547& 30.525& 30.545\\
\hline
$\delta\sigma_h$& 0.034& 0.034& 0.067& 0.034& 0.087& 0.067\\
\hline
\end{tabular}

\begin{tabular}{|c|c|c|c|c|}
\hline
$\sqrt{s}$ & 92.983& 93.035\\ \hline
$\sigma_h$  & 14.231& 13.91\\ \hline
$\delta\sigma_h$ & 0.046& 0.053\\ \hline
\end{tabular}

\bigskip

\begin{center}
Table 1.4 OPAL
\end{center}

\begin{tabular}{|c|c|c|c|c|c|c|c|}
\hline
$\sqrt{s}$& 89.4415& 89.45& 91.207& 91.222& 91.285& 92.973& 93.035 \\ \hline
$\sigma_h$& 9.980& 10.044& 30.445& 30.46& 30.64& 14.27& 13.85\\ \hline
$\delta\sigma_h$& 0.044& 0.034& 0.053& 0.025& 0.098& 0.046& 0.046\\ \hline
\end{tabular}

\end{document}